\documentclass[prd,prepint,superscriptaddress,notitlepage,showpacs,nofootinbib,longbibliography]{revtex4-1}

\usepackage{graphicx}
\usepackage[utf8]{inputenc} 
\usepackage{placeins}
\usepackage{amsmath}
\usepackage{amssymb}
\usepackage{float}
\usepackage{xcolor,color,colortbl}
\usepackage{comment}
\usepackage{multirow}
\usepackage{color}
\usepackage{soul}

\def\beq{\begin{equation}}
\def\eeq{\end{equation}}
\def\bea{\begin{eqnarray}}
\def\eea{\end{eqnarray}}

\usepackage{amsbsy}
\usepackage{bm}
\usepackage{color}
\usepackage{fancyhdr}
\usepackage[pdftex]{epsfig}
\usepackage[colorlinks=true,linkcolor=blue,citecolor=purple,urlcolor=magenta,filecolor=blue]{hyperref}
\usepackage{slashed}

\begin{document}

\bigskip

\vspace{2cm}

\title{Singlet vector leptoquark model facing recent LHCb and BABAR measurements}
\vskip 6ex

\author{Cristian H. Garc\'{i}a-Duque}
\email{chgarcia@uniquindio.edu.co}
\thanks{(Corresponding author)}
\affiliation{Programa de F\'{i}sica, Universidad del Quind\'{i}o, Carrera 15 Calle 12 Norte, C\'{o}digo Postal 630004, Armenia, Colombia}
\affiliation{Doctorado en Ciencias, Universidad del Quind\'{i}o, Carrera 15 Calle 12 Norte, C\'{o}digo Postal 630004, Armenia, Colombia}
\author{J. M. Cabarcas}
\email{josecabarcas@usantotomas.edu.co}
\affiliation{Universidad Santo Tom\'as, Colombia}
\author{J. H. Mu\~{n}oz}
\email{jhmunoz@ut.edu.co}
\affiliation{Departamento de F\'{i}sica, Universidad del Tolima, C\'{o}digo Postal 730006299, Ibagu\'{e}, Colombia}
\author{N\'{e}stor Quintero}
\email{nestor.quintero01@usc.edu.co}
\affiliation{Facultad de Ciencias B\'{a}sicas, Universidad Santiago de Cali, Campus Pampalinda, Calle 5 No. 62-00, C\'{o}digo Postal 76001, Santiago de Cali, Colombia}
\author{Eduardo Rojas}
\email{eduro4000@gmail.com}
\affiliation{Departamento de Física, Universidad de Nari\~{n}o, A.A. 1175, San Juan de Pasto, Colombia}

 \bigskip

\begin{abstract}
Very recently the LHCb experiment released the first measurement of the ratio $R(\Lambda_c) = {\rm BR}(\Lambda_b \to \Lambda_c\tau\bar{\nu}_\tau)/{\rm BR}(\Lambda_b \to \Lambda_c\mu\bar{\nu}_\mu)$. Moreover, the BABAR experiment reported a new result of the leptonic decay ratio of Upsilon meson $\Upsilon(3S)$, namely, $R_{\Upsilon(3S)} = {\rm BR}(\Upsilon(3S) \to \tau^+\tau^-)/{\rm BR}(\Upsilon(3S) \to \mu^+\mu^-)$. Both measurements are below their corresponding Standard Model predictions (deficit), deviating by $\sim 1.1\sigma$ and $\sim 1.8\sigma$, respectively. Moreover, the LHCb recently presented the first search of the lepton flavor violating decay $B^0 \to K^{\ast 0}\mu^\pm\tau^\mp$. Motivated by these new data, in this work we study their impact on the phenomenology of the singlet vector leptoquark ($U_1$) model addressing the hints of lepton flavor universality violation in the semileptonic decays of $B$ mesons ($B$ meson anomalies), by carrying out a global fit analysis. In general, we found that a minimal version of the $U_1$ model with a mass of 1.8 TeV can successfully explain the $B$ meson anomalies, while being compatible with all other flavor observables and LHC bounds. Interestingly, our study shows that the new observables $R(\Lambda_c)$ and $R_{\Upsilon(3S)}$ generate strong tension, leading to non-trivial effects on the global fit. Future improvements at the LHCb and Belle II experiments would help to understand their complementarity. Moreover, we also analyze the impact of the expected sensitivity on flavor observables at Belle II to provide a further test of the $U_1$ model. Finally, we study the minimal assumptions under which the $U_1$ model could, in addition, provide a combined explanation of the anomalous magnetic moment of the muon.
\end{abstract}

\maketitle


\section{Introduction}

The Standard Model (SM) of particle physics is until the present, the best-known theory for describing the dynamics of the fundamental constituents of the universe, excluding gravity. Despite its success, there are still open questions that are not answered by the SM, such as the number of families, neutrino masses, dark matter candidates, among others, which lead us to think that it corresponds to a low-energy effective theory of a more fundamental one. In the same route, in the SM the lepton flavor universality (LFU) states that in weak decays, there is no preference among the three lepton flavors, several experiments have looked for evidence of LFU-violation (LFUV) and thus for hints or signatures of new physics (NP). Particularly, during the last decade, there has been an accumulation of experimental results regarding \textit{B} meson transitions in tension with the SM predictions, namely, the charged and neutral \textit{B}-anomalies associated with $b \to c \tau \bar{\nu}_{\tau}$~\cite{Lees:2012xj,Lees:2013uzd,Huschle:2015rga,Sato:2016svk,Hirose:2017vbz,Aaij:2015yra,
Aaij:2017deq,Aaij:2017uff,Belle:2019rba,Hirose:2017dxl,Hirose:2016wfn,Abdesselam:2019wbt,HFLAV:2022pwe,
HFLAVsummer,Aaij:2017tyk} and $b \to s \mu^+ \mu^- $~\cite{Aaij:2014ora,Aaij:2019wad,LHCb:2021trn,Aaij:2017vbb,Abdesselam:2019lab,Abdesselam:2019wac,
LHCb:2021lvy,Aaij:2013qta,Aaij:2015oid,Aaij:2020nrf,Aaij:2013aln,Aaij:2015esa,Aaij:2020ruw} transitions, respectively. For a recent review, see Ref.~\cite{London:2021lfn}. Such anomalies offer excellent scenarios to test some NP models in order to explain simultaneously these tensions.  \medskip

The most recent charged-current LFU test is the observable 
\begin{equation}
 R(\Lambda_c) =\frac{ {\rm BR}(\Lambda_b \to \Lambda_c \tau \bar{\nu}_\tau)}{{\rm BR}(\Lambda_b \to \Lambda_c \mu \bar{\nu}_\mu)},  
 \end{equation} 
measured by the LHCb experiment~\cite{LHCb:2022piu}, 
\begin{equation} \label{RLambda_c}
    R(\Lambda_c) =
\begin{cases} 
\text{LHCb:} \ 0.242 \pm 0.026 \pm 0.040 \pm 0.059 \text{~\cite{LHCb:2022piu}}, \\
\text{SM:} \ 0.324 \pm 0.004 \text{~\cite{Bernlochner:2018bfn}},
\end{cases}  
\end{equation}

\noindent where the experimental uncertainties are statistical, systematic, and due to the external branching ratio measurement $\Lambda_b \to \Lambda_c \mu\bar{\nu}_\mu$ from LEP data, respectively~\cite{LHCb:2022piu}. This measurement is $\sim 1.1 \sigma$ below its corresponding SM prediction (deficit)~\cite{LHCb:2022piu}. This means that the $\Lambda_b \to \Lambda_c \ell \bar{\nu}_\ell$ process has a preference to decay to muon over tau lepton, $R(\Lambda_c)_{\rm LHCb} < R(\Lambda_c)_{\rm SM}$. 
Very recently, in Ref.~\cite{Bernlochner:2022hyz} was pointed out that by normalizing the LHCb measurement of $\Lambda_b \to \Lambda_c \tau \bar{\nu}_\tau$ to the SM prediction for $\Lambda_b \to \Lambda_c \mu \bar{\nu}_\mu$ (rather than LEP measurement), it provides a more consistent comparison with the SM prediction for $R(\Lambda_c)$. From this study was obtained a value of $R(\Lambda_c) = 0.285 \pm 0.073$~\cite{Bernlochner:2022hyz} with a higher central value and in agreement with SM at the $0.53\sigma$ level. Nevertheless, this $R(\Lambda_c) $ value also shows a suppression respect to the SM.
Intriguingly, this behaviour of $R(\Lambda_c)$ is contrary to the other $b \to c \tau \bar{\nu}_{\tau}$ observables such as the well known $R(D^{(\ast)})$ anomalies,
\begin{equation}
 R(D^{(\ast)}) =\frac{ {\rm BR}(B \to D^{(*)} \tau \bar{\nu}_\tau)}{{\rm BR}(B \to D^{(*)} \ell \bar{\nu}_\ell)}, \quad (\ell=\mu , e),
 \end{equation}

\noindent which world averages values reported by the Heavy Flavor Averaging Group (HFLAV)~\cite{HFLAV:2022pwe,HFLAVsummer}
\begin{eqnarray}
    R(D) &=& \label{RD}
\begin{cases}
\text{HFLAV:} \ 0.339 \pm 0.030, \\
\text{SM:} \ 0.298 \pm 0.004,
\end{cases}  \\
    R(D^\ast) &=&\label{RDstar}
\begin{cases}
\text{HFLAV:} \ 0.295\pm 0.014, \\
\text{SM:} \ 0.254  \pm 0.005,
\end{cases}  
\end{eqnarray}

\noindent exhibit a combined discrepancy of $\sim 3.3 \sigma$ above (excess) the SM~\cite{HFLAV:2022pwe,HFLAVsummer}, $R(D^{(\ast)})_{\rm HFLAV} > R(D^{(\ast)})_{\rm SM}$. The same is true for the ratio $R(J/\psi) = {\rm BR}(B_c \to J/\psi \tau \bar{\nu}_\tau)/{\rm BR}(B_c \to J/\psi\mu \bar{\nu}_{\mu})$~\cite{Aaij:2017tyk}, the $\tau$ lepton polarization $P_\tau(D^\ast)$~\cite{Hirose:2017dxl,Hirose:2016wfn} and the longitudinal polarization of the $D^*$ meson $F_L(D^\ast)$~\cite{Abdesselam:2019wbt} related with the channel $\bar{B} \to D^\ast \tau \bar{\nu}_\tau$, and the inclusive decay ratio $R(X_c) = {\rm BR}(B \to X_c\tau\bar{\nu}_\tau)/{\rm BR}(B \to X_c\mu\bar{\nu}_\mu)$~\cite{Kamali:2018bdp}, which also show a tension above the SM predictions~\cite{Kamali:2018bdp,Harrison:2020nrv,Tanaka:2012nw,Alok:2016qyh}.\medskip

Furthermore, it has been shown that the NP left-handed vector operator that explains  $b \to c \tau \bar{\nu}_\tau$ data, also generates effects on the neutral transition $b \bar{b} \to \tau^+ \tau^-$~\cite{Aloni:2017eny,Garcia-Duque:2021qmg}. The leptonic decay ratio of Upsilon mesons $\Upsilon (nS)$ ($n=1,2,3$) defined as 
\begin{equation} \label{R_Upsilon}
 R_{\Upsilon (nS)}  =\frac{{\rm BR}(\Upsilon (nS) \to  \tau^+{\tau^-})}{{\rm BR}(\Upsilon (nS) \to  \ell^+ {\ell^-})}, \quad (\ell=\mu , e),
 \end{equation} 

\noindent provides a very clean test of LFU~\cite{Aloni:2017eny}. In Table~\ref{TableRUpsilon} we summarize the current experimental measurements reported by BABAR and CLEO~\cite{delAmoSanchez:2010bt,Besson:2006gj,Lees:2020kom}, and the SM predictions (with an uncertainty typically of the order $\sim \mathcal{O}(10^{-5})$)~\cite{Aloni:2017eny}. These measurements are in good agreement with the SM estimations, except for the recent BABAR measurement on $R_{\Upsilon(3S)}$ that shows a tension at the $1.8\sigma$ level~\cite{Lees:2020kom}. In addition, the $R_{\Upsilon(3S)}$ average also deviates at the $1.7\sigma$ level with respect to the SM prediction~\cite{Garcia-Duque:2021qmg}, showing a deficit ($R_{\Upsilon(3S)}^{\rm Ave} < R_{\Upsilon(3S)}^{\rm SM}$). \medskip

\begin{table}[!t]
\centering
\renewcommand{\arraystretch}{1.2}
\renewcommand{\arrayrulewidth}{0.8pt}
\caption{\small Experimental status and SM predictions of the ratios $R_{\Upsilon(nS)}$ ($n=1,2,3$).}
\label{TableRUpsilon}	
\begin{tabular}{ccc}
\hline\hline
Ratio & Exp. measurement & SM prediction~\cite{Aloni:2017eny} \\
\hline
$R_{\Upsilon(1S)}$ & $1.005 \pm 0.013 \pm 0.022$ \ (BABAR~\cite{delAmoSanchez:2010bt}) & $0.9924$  \\
$R_{\Upsilon(2S)}$ & $1.04 \pm 0.04 \pm 0.05$ \ (CLEO~\cite{Besson:2006gj}) & $0.9940$ \\
$R_{\Upsilon(3S)}$ & $1.05 \pm 0.08 \pm 0.05$ \ (CLEO~\cite{Besson:2006gj}) & $0.9948$ \\
                                & $0.966 \pm 0.008 \pm 0.014$ \ (BABAR~\cite{Lees:2020kom}) &  \\
                                & $0.968 \pm 0.016$ \ (Average~\cite{Garcia-Duque:2021qmg}) & \\
\hline\hline
\end{tabular} 
\end{table}

The interesting fact that the $b \to c \tau \bar{\nu}_{\tau}$ data reflect an excess with respect to the SM, except for the ratio $R(\Lambda_c)$ reported by LHCb~\cite{LHCb:2022piu}, and that the ratio $R_{\Upsilon (3S)}$ also shows a deficit~\cite{Lees:2020kom}, raises the question: How does it impact a global phenomenological analysis, in any model beyond the SM used to explain simultaneously both the charged- and neutral-current $B$ meson anomalies? In this work, we study such an impact in one of the most promising NP models for addressing  these flavor anomalies, the well-known singlet vector leptoquark $U_1 \equiv U_1 \sim (\textbf{3},\textbf{1},2/3)$~\cite{London:2021lfn,Sakaki:2013bfa,Freytsis:2015qca,Bhattacharya:2016mcc,Kumar:2018kmr,Calibbi:2015kma,Feruglio:2018fxo,Angelescu:2018tyl,Angelescu:2021lln,Iguro:2018vqb,Baker:2019sli,
Cornella:2019hct,Cornella:2021sby,Hati:2020cyn,Hati:2019ufv,Iguro:2020keo,Faroughy:2016osc,Buttazzo:2017ixm,
Hati:2019ufv,Bordone:2017bld,Bordone:2018nbg,Greljo:2018tuh,Calibbi:2017qbu,Blanke:2018sro,
Bernigaud:2021fwn,Crivellin:2018yvo,Assad:2017iib,Fornal:2018dqn,FernandezNavarro:2022gst}, which is a $SU(3)_c$ triplet, $SU(2)_L$ singlet, and hypercharge 2/3. In the existing literature, two general approaches for the description of the $U_1$ model have been done. The first one starts from a phenomenological approach which introduces particular textures for the couplings of the leptoquark to the left-handed (LH) and right-handed (RH) SM fermions. However, in this procedure there is a limitation of the estimation of some one loop leptoquark contributions to low-energy processes (such as $\tau \to \mu\gamma$, $B_s - \bar{B}_s$ mixing). This reason motivates the second approach which is based on the construction of a complete ultra-violet (UV) model to achieve the desire pattern of couplings, where the introduction of additional flavor symmetries is often required, new vector-like families are also needed, as well as extra scalar fields to achieve properly the symmetry breaking mechanism, see e.g. Refs.~\cite{Cornella:2019hct,Cornella:2021sby,Hati:2019ufv,Bordone:2017bld,Bordone:2018nbg,Buttazzo:2017ixm,
Greljo:2018tuh,Hati:2020cyn,Calibbi:2017qbu,Blanke:2018sro,Assad:2017iib,Fornal:2018dqn,
FernandezNavarro:2022gst}. 
In this study, we will work under the phenomenological approach based on the minimal setup of couplings (flavor-dependent) between $U_1$ and
LH fermions of the SM with vanishing RH quark-lepton couplings, without specifying the underlying theory. This is the so-called \textit{minimal $U_1$ model} and we will closely follow the notation of Refs.~\cite{Angelescu:2018tyl,Angelescu:2021lln}. Interestingly, very recently in Ref.~\cite{FernandezNavarro:2022gst} was proposed a complete UV model (the twin Pati-Salam model) in which purely LH $U_1$ couplings are naturally predicted with fulfilled conditions. Moreover, it was shown in Ref.~\cite{Belanger:2022kvj} that the minimal $U_1$ model can be extended with a scalar dark matter (DM) candidate that couples to the $U_1$ to explain different DM observables. \medskip

Before the result of LHCb on $R(\Lambda_c)$~\cite{LHCb:2022piu}, a recent analysis of the minimal $U_1$ model presented in Ref.~\cite{Angelescu:2021lln} predicted an increasing of $R(\Lambda_c)$ with respect to the SM, $R(\Lambda_c)/R(\Lambda_c)_{\rm SM}  = 1.15 \pm 0.10$, which is clearly in contradiction with the LHCb measurement that exhibit a suppression, $R(\Lambda_c)_{\rm LHCb}/R(\Lambda_c)_{\rm SM} = 0.75 \pm 0.23$~\cite{LHCb:2022piu}. In addition, in Refs.~\cite{Blanke:2018yud,Blanke:2019qrx} were obtained a sum rule relating the ratios $R(D)$ and $R(D^\ast)$ with $R(\Lambda_c)$ which holds for any NP scenario and provides an important cross-check of the experimental values of $R(D^{(\ast)})$. The sum rule predicted an enhancement of $R(\Lambda_c)$ with respect to its SM value, $R(\Lambda_c)/R(\Lambda_c)_{\rm SM} = 1.15 \pm 0.04$~\cite{Blanke:2019qrx}. This shows that the suppressed value of $R(\Lambda_c)$ obtained by LHCb is (again) inconsistent with the excess found in $R(D^{(\ast)})$. To sum up, the experimental result of LHCb is challenging the theoretical predictions. \medskip

The main goal of this work is to perform a global fit analysis of the parametric space of the minimal $U_1$ model by considering the impact of the new measurements $R(\Lambda_c)$ from LHCb~\cite{LHCb:2022piu} and $R_{\Upsilon(3S)}$ from BABAR~\cite{Lees:2020kom} to the $b \to c \tau \bar{\nu}_\tau$ and $b \to s \mu^+\mu^-$ ($C^{bs\mu\mu}_{9} = - C^{bs\mu\mu}_{10}$) data. We take into account LHC constraints to the model~\cite{Angelescu:2021lln} and several low-energy processes that are induced at the tree-level such as, lepton flavor violating (LFV) decays ($B \to K^{(\ast)} \mu^\pm \tau^\mp$, $B_s \to \mu^\pm \tau^\mp$, $\tau \to \mu\phi$, $\Upsilon(nS) \to \mu^\pm \tau^\mp$) and rare $B$ decays ($B \to K \tau^+ \tau^-, B_s \to \tau^+ \tau^-$). Furthermore, we also analyze the expected sensitivity on flavor observables at Belle II (for an integrated luminosity of $50 \ {\rm ab}^{-1}$~\cite{Kou:2018nap}) to provide further test of the $U_1$ model. At the end, we analyze the additional assumptions under which the minimal $U_1$ model can in addition provide a combined explanation of the anomalous magnetic moment of the muon $(g-2)_\mu$, without affecting the parametric space addressing the $B$ meson anomalies. Some previous works have studied the possibility of a common solution of $(g-2)_\mu$ and the $B$ meson anomalies~\cite{Altmannshofer:2020ywf,Du:2021zkq,Ban:2021tos}, in which the $U_1$ leptoquark couples to both LH and RH fermions. We will show that by allowing only one RH coupling different from zero, it is possible to get a combined explanation within the minimal $U_1$ model.\medskip

We structured this work as follows: In section~\ref{model} we give a brief description of the $U_1$ vector leptoquark model. In section~\ref{observables} we present the various relevant processes to which the minimal $U_1$ model contributes ($b \to c \tau \bar{\nu}_\tau$, $b \to s \mu^+\mu^-$, $\Upsilon$ decays, LFV and rare decays). We then carry out our phenomenological analysis of the allowed parametric space in section~\ref{analysis}. In section \ref{gminus2} we extent economically the model (by including a single right-handed coupling) to adjust $(g-2)_\mu$ data. Our main conclusions are presented in section~\ref{Conclusion}. \medskip

\section{Singlet vector leptoquark model: $U_1$}
 \label{model}

The interaction of the $SU(2)_L$ singlet vector leptoquark (LQ) $U_1$ with the SM fermions can be written as~\cite{Angelescu:2018tyl,Angelescu:2021lln}
\begin{equation}
\Delta\mathcal{L}_{U_1} = (x_L^{ij} \ \bar{Q}_{iL} \gamma_\mu L_{jL} + x_R^{ij} \ \bar{d}_{iR} \gamma_\mu \ell_{jR}) U_1^\mu ,
\end{equation}

\noindent where the LH and RH quark-lepton flavor couplings $x_L$ and $x_R$ are (in general) complex $3\times 3$ matrices, $Q_{L}$ and $L_{L}$ are the LH quark and lepton doublets defined as
\begin{equation}
Q_{iL} = \begin{pmatrix}
V_{ki}^\dagger u_{kL} \\
d_{iL}
\end{pmatrix}, \hspace{1cm} 
L_{jL} = \begin{pmatrix}
\nu_{jL} \\
\ell_{jL}
\end{pmatrix}, 
\end{equation}

\noindent respectively, with $V$ denoting the Cabibbo-Kobayashi-Maskawa (CKM) matrix; and $\ell_R$ and $d_R$ are the RH charged leptons and down-type quarks singlets. We will consider a minimalistic flavor structure of the LH coupling matrix $x_L$ and assume
\begin{equation} \label{FS}
x_L = \begin{pmatrix}
0 & 0 & 0 \\
0 & x_L^{s\mu} & x_L^{s\tau} \\
0 & x_L^{b\mu} & x_L^{b\tau} \\
\end{pmatrix} , 
\end{equation}

\noindent neglecting couplings to the first generation of quarks and leptons. While for the RH sector, we will assume vanishing couplings  $(x_R =0)$.\footnote{For analyses taking into account non-vanishing RH couplings, see, e.g. Refs.~\cite{Cornella:2019hct,Cornella:2021sby}.} This is the so-called \textit{minimal $U_1$ model}~\cite{Angelescu:2018tyl,Angelescu:2021lln}. In this work, we will take these flavor-dependent couplings (involving only second and third generations) to be real. Let us stress that for the phenomenological analysis of the minimal $U_1$ model, we will assume that RH couplings are zero; however, by allowing only one RH coupling to bottom-quark and muon different from zero ($x_R^{b\mu} \neq 0$), it is possible to obtain an enhanced effect on the anomalous magnetic moment of the muon, as we will discuss in Sec.~\ref{gminus2}. \medskip

After integrating out the Lagrangian $\Delta\mathcal{L}_{U_1}$, the flavor structure given by Eq.~\eqref{FS} generates tree-level contributions to neutral-current $b \to s \mu^+\mu^-$ and charged-current $b \to c \tau^- \bar{\nu}_\tau $ processes. Moreover, this $U_1$ model allows to induce other flavor observables, such as LFV decays ($B \to K^{(\ast)} \mu^\pm \tau^\mp$, $B_s \to \mu^\pm \tau^\mp$, $\tau \to \mu\phi$, $\Upsilon(nS) \to \mu^\pm \tau^\mp$), and rare $B$ decays ($B \to K \tau^+ \tau^-, B_s \to \tau^+ \tau^-$). In addition, this scenario gives rise to the neutral-current $b\bar{b} \to \tau^+\tau^-$ transiton, thus generating effects on the leptonic decay ratio of Upsilon mesons $R_{\Upsilon(nS)}$, see Eq.~\eqref{R_Upsilon}. In most of the recent studies of the $U_1$ model~\cite{Angelescu:2018tyl,Angelescu:2021lln,Cornella:2019hct,Cornella:2021sby}, the implications of these bottomonium observables are not usually taken into account. We will properly include them in our study.

\section{Flavor observables} \label{observables}

In this section, we present the various relevant processes to which the minimal $U_1$ model contributes and summarize all the experimental constraints. It is well known that the singlet LQ $U_1$ does not generates tree-level contributions to the FCNC transition $b \to s \nu \bar{\nu}$ ($B \to K^{(\ast)} \nu \bar{\nu}$ processes)~\cite{Kumar:2018kmr}. For this reason we will not include it in our analysis. In addition, within our phenomenological approach of the minimal $U_1$ model, we will not consider the one-loop induced processes $\tau \to \mu\gamma$ and $B_s - \bar{B}_s$ mixing which might be sensitive (model dependent) to the features of a specific UV completion~\cite{Angelescu:2018tyl,Angelescu:2021lln}.

\subsection{Charged-current $b \to c \tau^- \bar{\nu}_\tau$ processes}

The effective Hamiltonian responsible for the charged-current $b \to c \tau \bar{\nu}_\tau$ transition is given by
\bea\label{Heff_CC}
 \mathcal{H}_{\rm eff}(b \to c \tau \bar{\nu}_{\tau}) &=& \frac{4 G_F}{\sqrt{2}} V_{cb}	 \Big[(1+ C_{V}^{bc\tau\nu_\tau})(\bar{c} \gamma_\mu P_L b) (\bar{\tau} \gamma^\mu P_L \nu_{\tau})  \Big], 
\eea
\noindent where $V_{cb}$ denotes the Cabbibo-Kobayashi-Maskawa (CKM) matrix element, $G_F$ is the Fermi coupling constant, and $C_{V}^{bc\tau\nu_\tau}$ is the Wilson coefficient (WC) which in the $U_1$ LQ scenario read as~\cite{Angelescu:2018tyl,Angelescu:2021lln}
\begin{eqnarray}
C_{V}^{bc\tau\nu_\tau} &=& \frac{\sqrt{2}}{4 G_F V_{cb} M_{U_1}^2} (Vx_L)^{c\tau}(x_L^{b\tau})^\ast , \\
&=&  \frac{\sqrt{2}}{4 G_F  M_{U_1}^2} \Big[\vert x_L^{b\tau} \vert^2 + \frac{V_{cs}}{V_{cb}} x_L^{s\tau} (x_L^{b\tau})^\ast \Big],
\end{eqnarray}

\noindent with $M_{U_1}$ the vector $U_1$ mass. The contribution of $U_1$ model leads to a re-scaling of all $b \to c \tau \bar{\nu}_\tau$ observables, namely 
\begin{eqnarray}
R(H) &=& R(H)_{\rm SM} \big| 1 + C_{V}^{bc\tau\nu_\tau} \big|^2,  \ \ \ (\text{with} \ H=D,D^\ast,J/\psi,\Lambda_c) \label{RH} \\
F_L(D^*) &=&  F_L(D^*)_{\rm SM} \ \Big( \frac{R(D^*)}{R(D^*)_{\rm SM}}\Big)^{-1}   \big| 1 + C_{V}^{bc\tau\nu_\tau}\big|^2 , \label{FLD} \\
P_\tau(D^*) &=&  P_\tau(D^*)_{\rm SM} \ \Big( \frac{R(D^*)}{R(D^*)_{\rm SM}}\Big)^{-1} \big\vert 1 + C_{V}^{bc\tau\nu_\tau} \big\vert^2 \ , \label{PTAU} \\
R(X_c) &=& R(X_c)_{\rm SM} \Big( 1+ 2.294 \ {\rm Re}(C_{V}^{bc\tau\nu_\tau}) + 1.147 \big\vert C_{V}^{bc\tau\nu_\tau} \big \vert^2 \Big ), \label{R_Xc} \\
{\rm BR}(B_c^- \to \tau^- \bar{\nu}_{\tau}) &=& {\rm BR}(B_c^- \to \tau^- \bar{\nu}_{\tau})_{\text{SM}}   \big\vert 1+C_{V}^{bc\tau\nu_\tau} \big\vert^2 . \label{BRBc_modified}
\end{eqnarray}

\noindent The experimental measurements and SM predictions of $R(D)$, $R(D^\ast)$, and $R(\Lambda_c)$ are given by Eqs.~\eqref{RD},~\eqref{RDstar}, and~\eqref{RLambda_c}, respectively. For the other $b \to c \tau \bar{\nu}_\tau$ observables, we collect both the experimental and theoretical values in Table~\ref{Table:2}. The tauonic channel $B_c^- \to \tau^- \bar{\nu}_{\tau}$  has not been measured yet, but indirect constraints on ${\rm BR}(B_c^- \to \tau^- \bar{\nu}_{\tau})$ have been imposed using the lifetime of $B_c$ $(< 30\%)$~\cite{Alonso:2016oyd} and from LEP data at the $Z$ peak $(< 10\%)$~\cite{Akeroyd:2017mhr}. In addition, a conservative bound of ${\rm BR}(B_c^- \to \tau^- \bar{\nu}_{\tau}) \lesssim 40\%$ has also been obtained in~\cite{Celis:2016azn}. In further analysis we will use the bound of $10\%$~\cite{Akeroyd:2017mhr}.

\begin{table}[!t]
\centering
\renewcommand{\arraystretch}{1.4}
\renewcommand{\arrayrulewidth}{0.8pt}
\caption{\small Experimental measurements and SM predictions on other $b \to c \tau \bar{\nu}_\tau$ observables.}
\begin{tabular}{ccc}
\hline
Observable & Expt. measurement & SM prediction \\
\hline
$R(J/\psi)$ & $0.71 \pm 0.17 \pm 0.18$~\cite{Aaij:2017tyk} & 0.2582 $\pm$ 0.0038~\cite{Harrison:2020nrv}  \\
$P_\tau(D^\ast)$ & $- 0.38 \pm 0.51 ^{+0.21}_{-0.16}$~\cite{Hirose:2017dxl,Hirose:2016wfn} &  $-0.497 \pm 0.013$~\cite{Tanaka:2012nw} \\
$F_L(D^\ast)$ & $0.60 \pm 0.08 \pm 0.035$~\cite{Abdesselam:2019wbt} & $0.46 \pm 0.04$~\cite{Alok:2016qyh}  \\
$R(X_c)$ & 0.223 $\pm$ 0.030~\cite{Kamali:2018bdp} & 0.216 $\pm$ 0.003~\cite{Kamali:2018bdp} \\
\hline
\end{tabular} \label{Table:2}
\end{table}

Concerning to the LFU ratio, $R_D^{\mu/e} \equiv {\rm BR}(B\to D\mu \bar{\nu}_\mu)/{\rm BR}(B\to De \bar{\nu}_e)$, the SM estimation~\cite{Becirevic:2020rzi} is in excellent agreement with the experimental value reported by Belle~\cite{Glattauer:2015teq}, namely
\begin{equation} \label{RD_mue}
    R_D^{\mu/e} =
\begin{cases} 
\text{Belle:} \ 0.995 \pm 0.022 \pm 0.039 \text{~\cite{Glattauer:2015teq}}, \\
\text{SM:} \ 0.9960\pm 0.0002 \text{~\cite{Becirevic:2020rzi}}.
\end{cases}  
\end{equation}

\noindent The $U_1$ leptoquark modifies this ratio as 
\begin{equation}
R_D^{\mu/e} = [R_D^{\mu/e}]_{\rm SM}\big|1+C_{V}^{bc\mu\nu_\mu} \big|^2.
\end{equation}

\noindent with 
\begin{equation}
C_{V}^{bc\mu\nu_\mu} =  \frac{\sqrt{2}}{4 G_F  M_{U_1}^2} \Big[\vert x_L^{b\mu} \vert^2 + \frac{V_{cs}}{V_{cb}} x_L^{s\mu} (x_L^{b\mu})^\ast \Big].
\end{equation}

Last but no least, the minimal $U_1$ model can also induce NP contributions in the charged-current transition $b \to u\tau\bar{\nu}_\tau$, such is the case of the leptonic decay $B\to \tau\bar{\nu}_\tau$~\cite{Angelescu:2018tyl}. Its branching fraction can be rescaled as
\begin{equation}\label{BRBu}
{\rm BR}(B^- \to \tau^- \bar{\nu}_{\tau}) = {\rm BR}(B^- \to \tau^- \bar{\nu}_{\tau})_{\text{SM}}   \big\vert 1+C_{V}^{bu\tau\nu_\tau} \big\vert^2 ,
\end{equation}
where
\begin{equation}
C_{V}^{bu\tau\nu_\tau} =  \frac{\sqrt{2}}{4 G_F  M_{U_1}^2} \Big[\vert x_L^{b\tau} \vert^2 + \frac{V_{us}}{V_{ub}} x_L^{s\tau} (x_L^{b\tau})^\ast \Big],
\end{equation}

\noindent with $V_{ub}$ and $V_{us}$ denoting the CKM matrix elements involved. The current experimental value reported by the Particle Data Group (PDG)~\cite{PDG2020} and its corresponding SM estimation,
\begin{equation} \label{Btotaunu}
 {\rm BR}(B^- \to \tau^- \bar{\nu}_{\tau}) =
\begin{cases} 
\text{PDG:} \ (1.09 \pm 0.24)\times 10^{-4} \text{~\cite{PDG2020}}, \\
\text{SM:} \ (0.989 \pm 0.013)\times 10^{-4},
\end{cases}  
\end{equation}

\noindent respectively, reflects an excellent agreement ($0.4\sigma$). The theoretical value was obtained by using $f_B=(190.0\pm1.3)$ MeV and $V_{ub} = (3.94 \pm 0.36)\times 10^{-3}$ from PDG~\cite{PDG2020}. 

\subsection{Neutral-current $b \to s \mu^+\mu^-$ processes}

The  $U_1$ vector LQ contributes at the tree-level to $b \to s\mu^+\mu^-$ transitions via the effective Hamiltonian~\cite{Angelescu:2018tyl,Angelescu:2021lln,Bhattacharya:2016mcc,Kumar:2018kmr}
\begin{equation}
\mathcal{H}_{\rm eff}(b \to s \mu^+\mu^-) = -\dfrac{\alpha_{\rm em} G_F}{\sqrt{2}\pi} V_{tb} V_{ts}^\ast \big[C^{bs\mu\mu}_9 \mathcal{O}_{9} + C^{bs\mu\mu}_{10} \mathcal{O}_{10} \big],
\end{equation}

\noindent where $\alpha_{\rm em}$ is the fine-constant structure, $\mathcal{O}_9 = (\bar{s}P_L \gamma_\beta b) (\bar{\mu}\gamma^\beta \mu)$, $\mathcal{O}_{10} = (\bar{s}P_L \gamma_\beta b) (\bar{\mu} \gamma^\beta \gamma_5 \mu)$, and the WCs read as
\begin{equation}
C^{bs\mu\mu}_{9} = - C^{bs\mu\mu}_{10} = - \frac{\pi}{\sqrt{2} G_F \alpha_{\rm em} V_{tb} V_{ts}^\ast} \frac{x_L^{s\mu}(x_L^{b\mu})^\ast}{M_{U_1}^2}.
\end{equation}

\noindent Regarding the $b \to s\mu^+\mu^-$ data~\cite{Aaij:2014ora,Aaij:2019wad,LHCb:2021trn,Aaij:2017vbb,Abdesselam:2019lab,Abdesselam:2019wac,
LHCb:2021lvy,Aaij:2013qta,Aaij:2015oid,Aaij:2020nrf,Aaij:2013aln,Aaij:2015esa,Aaij:2020ruw},  the largest deviations of $R_{K^{(\ast)}} = {\rm BR}(B \to K^{(\ast)}\mu^+\mu^-)/{\rm BR}(B \to K^{(\ast)}e^+e^-)$, have been observed by the LHCb, hinting toward LFU violation. Moreover, there are some additional anomalous observables such as the $B_s \to  \phi  \mu^+\mu^-$ decay rate, and angular observables and differential branching fractions related with $B \to K^\ast  \mu^+\mu^-$ decay. 
These data can be explained if there is new physics (NP) effects in $b \to s \mu^+\mu^-$, i.e., the hypothesis that NP couples selectively to the muons. Several global fit analyses taking into account the most recent $b \to s \mu^+\mu^-$ observables have been performed in the literature~\cite{Aebischer:2019mlg,Altmannshofer:2021qrr, Alguero:2021anc,Alguero:2019ptt,Geng:2021nhg,Hurth:2021nsi,Angelescu:2021lln,Carvunis:2021jga,London:2021lfn}, showing that the operators $\mathcal{O}_{9(10)}$ provides an excellent description of the data~\cite{Aebischer:2019mlg,Altmannshofer:2021qrr, Alguero:2021anc,Alguero:2019ptt,Geng:2021nhg,Hurth:2021nsi,Angelescu:2021lln,Carvunis:2021jga}. 
We will adopt the results of the recent global analysis performed in Ref.~\cite{Altmannshofer:2021qrr}. According to the fit~\cite{Altmannshofer:2021qrr}, the allowed $1\sigma$ solution to the WC is 
\begin{equation} \label{C9}
C^{bs\mu\mu}_{9} = - C^{bs\mu\mu}_{10} \in [-0.46,-0.32].
\end{equation}

\noindent Thus, one obtains 
\begin{equation}
 -\frac{x_L^{s\mu}(x_L^{b\mu})^\ast }{M_{U_1}^2} \in [4.8, 6.9] \times 10^{-4} \ {\rm TeV}^{-2}.
\end{equation}

\subsection{Upsilon decay ratio $R_{\Upsilon(nS)}$ ($n=1,2,3$)}

The tree-level $U_1$ LQ effects on the leptonic decay ratio of Upsilon mesons $R_{\Upsilon(nS)}$ (Eq.~\eqref{R_Upsilon}) can be written as~\cite{Aloni:2017eny}
\begin{equation}
R_{\Upsilon(nS)}= \frac{(1-4x_\tau^2)^{1/2}}{\vert A_V^{\rm SM} \vert^2} \Big[ \vert A_V^{b\tau} \vert^2 (1+2x_\tau^2) + \vert B_V^{b\tau} \vert^2 (1- 4x_\tau^2) \Big],
\end{equation}

\noindent with $x_\tau = m_\tau/m_{\Upsilon(nS)}$, $\vert A_V^{\rm SM} \vert = - 4\pi\alpha_{\rm em} Q_b$ ($Q_b = -1/3$), and 
\begin{eqnarray}
A_V^{b\tau} &=& - 4\pi\alpha_{\rm em} Q_b + \frac{m_{\Upsilon(nS)}^2}{4} \Big( -\frac{\vert x_L^{b\tau}\vert^2}{M_{U_1}^2} \Big), \\
B_V^{b\tau} &=&- \frac{m_{\Upsilon(nS)}^2}{2} \Big( -\frac{\vert x_L^{b\tau}\vert^2}{M_{U_1}^2} \Big).
\end{eqnarray}

\subsection{LFV decays}
This section is dedicated to the study of LFV decay channels of the $B$ meson, $\tau$ lepton and Upsilon mesons $\Upsilon(nS)$ ($n=1,2,3$), which occur at the tree-level due to the exchange of the $U_1$ vector LQ. The effective Hamiltonian for the LFV transitions $b \to s \mu^\mp \tau^\pm$, $\tau^- \to \mu^- s\bar{s}$, and $b\bar{b} \to \mu^\mp \tau^\pm$ can be generically written as~\cite{Kumar:2018kmr}
\begin{equation}
 \mathcal{H}_{\rm eff}^{\rm LFV}  = -\dfrac{\alpha_{\rm em}G_F}{\sqrt{2}\pi} V_{tb} V_{ts}^\ast \big[C^{qq^\prime\mu\tau}_9 (\bar{q}^\prime P_L \gamma_\beta q) (\bar{\mu}\gamma^\beta \tau) + C^{qq^\prime\mu\tau}_{10} (\bar{q}^\prime P_L \gamma_\beta q) (\bar{\mu} \gamma^\beta \gamma_5 \tau) \big] \ \ (q^{(\prime)}= b,s),
\end{equation}

\noindent where the WCs are
\begin{eqnarray}
C^{bs\mu\tau}_{9} = - C^{bs\mu\tau}_{10} &=& - \frac{\pi}{\sqrt{2} G_F \alpha_{\rm em} V_{tb} V_{ts}^\ast} \frac{x_L^{s\tau}(x_L^{b\mu})^\ast}{M_{U_1}^2}, \\
C^{ss\mu\tau}_{9} = - C^{ss\mu\tau}_{10} &=& - \frac{\pi}{\sqrt{2} G_F \alpha_{\rm em} V_{tb} V_{ts}^\ast} \frac{x_L^{s\tau}(x_L^{s\mu})^\ast}{M_{U_1}^2}, \\
C^{bb\mu\tau}_{9} = - C^{bb\mu\tau}_{10} &=& - \frac{\pi}{\sqrt{2} G_F \alpha_{\rm em} V_{tb} V_{ts}^\ast} \frac{x_L^{b\tau}(x_L^{b\mu})^\ast}{M_{U_1}^2},
\end{eqnarray}

\noindent respectively. This leads to the following processes $B \to K^{(\ast)} \mu^\pm \tau^\mp$, $B_s \to \mu^\pm \tau^\mp$, $\tau \to \mu \phi$, and $\Upsilon(nS) \to \mu^\pm \tau^\mp$. In Table~\ref{TableLFV} we list the current experimental upper limit (UL) on the branching ratios of these LFV decays~\cite{PDG2020,Aaij:2020mqb,LHCb:2022wrs,Aaij:2019okb}. These include the first search of $B^0 \to K^{\ast 0}\mu^\pm\tau^\mp$ recently performed by LHCb~\cite{LHCb:2022wrs}. We also show for some of these processes the Belle II experiment expected sensitivity for an integrated luminosity of 50 ab$^{-1}$~\cite{Kou:2018nap}.

\begin{table}[!t]
\centering
\renewcommand{\arraystretch}{1.2}
\renewcommand{\arrayrulewidth}{0.8pt}
\caption{\small Experimental status and Belle II future sensitivity of different LFV processes and rare $B$ decays.}
\label{TableLFV}	
\begin{tabular}{ccc}
\hline\hline
Channel & Current UL (at $90\%$ CL) & Belle II future sensitivity \\
\hline
$B^+ \to K^+\mu^+\tau^-$ & $4.5 \times 10^{-5}$ \ (PDG~\cite{PDG2020}) & $3.3 \times 10^{-6}$ \\
$B^+ \to K^+\mu^-\tau^+$ & $2.8 \times 10^{-5}$ \ (PDG~\cite{PDG2020}) & $3.3 \times 10^{-6}$ \\
                                               & $3.9 \times 10^{-5}$ \ (LHCb~\cite{Aaij:2020mqb})  &  \\
$B^0 \to K^{\ast 0}\mu^+\tau^-$ & $8.2 \times 10^{-6}$ \ (LHCb~\cite{LHCb:2022wrs}) & \\
$B^0 \to K^{\ast 0}\mu^-\tau^+$ & $1.0 \times 10^{-5}$ \ (LHCb~\cite{LHCb:2022wrs}) &  \\                                               
$B_s \to \mu^{\pm} \tau^{\mp}$ & $3.4 \times 10^{-5}$ \ (LHCb~\cite{Aaij:2019okb}) & \\
$\tau \to \mu \phi$ & $8.4 \times 10^{-8}$ \ (PDG~\cite{PDG2020}) & $\sim 2.0 \times 10^{-9}$ \\
$\Upsilon(1S) \to \mu^\pm \tau^\mp$ & $6 \times 10^{-6}$ \ (PDG~\cite{PDG2020})  & \\
$\Upsilon(2S) \to \mu^\pm \tau^\mp$ & $3.3 \times 10^{-6}$ \ (PDG~\cite{PDG2020})  & \\
$\Upsilon(3S) \to \mu^\pm \tau^\mp$ & $3.1 \times 10^{-6}$ \ (PDG~\cite{PDG2020})  & $\sim 10^{-7}$~\cite{Bhattacharya:2016mcc} \\
\hline
$B \to K \tau^+ \tau^-$ & $6.8 \times 10^{-3}$ \ (LHCb~\cite{Aaij:2017xqt}) & $8.1 \times 10^{-4}$ (for 5 ab$^{-1}$) \\
$B_s \to \tau^+ \tau^-$ & $2.25 \times 10^{-3}$ \ (PDG~\cite{PDG2020}) & $2.0 \times 10^{-6}$ \\
\hline\hline
\end{tabular} 
\end{table}


\subsubsection{$B \to K^{(\ast)} \mu^\pm \tau^\mp$ and $B_s \to \mu^\pm \tau^\mp$}

The branching ratio of the LFV decays $B^+ \to K^+\mu^+\tau^-$ and $B^0 \to K^{\ast 0}\mu^+\tau^-$ can be expressed as~\cite{Bhattacharya:2016mcc,Kumar:2018kmr,Calibbi:2015kma}
\begin{eqnarray}
{\rm BR}(B^+ \to K^+\mu^+\tau^-) &=& \big(a_{K} \vert C^{bs\mu\tau}_{9} \vert^2+ b_{K} \vert C^{bs\mu\tau}_{10} \vert^2 \big) \times 10^{-9} , \\
{\rm BR}(B^0 \to K^{\ast 0}\mu^+\tau^-) &=& \Big( (a_{K^\ast}+c_{K^\ast}) \vert C^{bs\mu\tau}_{9} \vert^2 + (b_{K^\ast}+d_{K^\ast}) \vert C^{bs\mu\tau}_{10} \vert^2 \Big) \times 10^{-9} ,
\end{eqnarray}

\noindent respectively, where $a_{K} = 9.6 \pm 1.0$, $b_{K} = 10.0 \pm 1.3$, $a_{K^\ast}=3.0\pm 0.8$, $b_{K^\ast}=2.7\pm 0.7$, $c_{K^\ast}=16.4\pm 2.1$, and $d_{K^\ast}=15.4\pm 1.9$ are  numerical coefficients that have been calculated using the $B\to K^{(\ast)}$ transitions form factors obtained from lattice QCD~\cite{Calibbi:2015kma}. The decay channel with final state $\mu^-\tau^+$ can be easily obtained by replacing $\mu \leftrightarrows \tau$. Let us notice that the LHCb limit on ${\rm BR}(B^+ \to K^+\mu^-\tau^+)$~\cite{Aaij:2020mqb} is comparable with the one quoted from PDG~\cite{PDG2020} (see Table~\ref{TableLFV}).\medskip

As for the LFV leptonic decay $B_s \to \mu^\pm \tau^\mp$, the branching ratio is written as~\cite{Calibbi:2015kma}
\begin{eqnarray}
{\rm BR}(B_s^0 \to \mu^\pm \tau^\mp) &=& \tau_{B_s}\frac{f_{B_s}^2 m_{B_s} m^2_\tau}{32\pi^3} \alpha_{\rm em}^2 G_F^2 \vert V_{tb} V_{ts}^\ast \vert^2 \Big(1-  \frac{m_\tau^2}{m_{B_s}^2} \Big)^2 \big(\vert C^{bs\mu\tau}_{9} \vert^2+  \vert C^{bs\mu\tau}_{10} \vert^2 \big), 
\end{eqnarray}

\noindent where $f_{B_s} = (230.3 \pm 1.3)$ MeV is the $B_s$ decay constant~\cite{HFLAV:2022pwe}. The last expression was obtained by using the limit $m_\tau \gg m_\mu$. 

\subsubsection{$\tau \to \mu \phi$}
For the LFV hadronic $\tau$ decay $\tau \to \mu \phi$ ($\tau \to \mu s\bar{s}$ transition), the branching ratio is computed as~\cite{Bhattacharya:2016mcc,Kumar:2018kmr}
\begin{equation}
{\rm BR}(\tau^- \to \mu^-\phi) =  \frac{f_\phi^2 m_\tau^3}{128\pi \Gamma_\tau} \Big(1+2 \frac{m_\phi^2}{m_\tau^2} \Big) \Big(1- \frac{m_\phi^2}{m_\tau^2} \Big)^2 \Big\vert \dfrac{x_L^{s\tau}(x_L^{s\mu})^\ast}{M_{U_1}^2} \Big\vert^2,
\end{equation}

\noindent where $m_\phi$ and  $f_\phi = (238 \pm 3)$ MeV~\cite{Kumar:2018kmr} are the $\phi$ meson mass and  decay constant, respectively. 

\subsubsection{$\Upsilon(nS) \to \mu^\pm \tau^\mp$}
The LFV leptonic $\Upsilon \equiv \Upsilon(nS) \ (n=1,2,3)$ decay is given by~\cite{Bhattacharya:2016mcc,Kumar:2018kmr}
\begin{equation}\label{upsilon}
{\rm BR}(\Upsilon \to \mu^\pm \tau^\mp) =  \frac{f_\Upsilon^2 m_\Upsilon^3}{48\pi \Gamma_\Upsilon} \Big(2+ \frac{m_\tau^2}{m_\Upsilon^2} \Big) \Big(1- \frac{m_\tau^2}{m_\Upsilon^2} \Big)^2 \Big\vert \dfrac{x_L^{b\tau}(x_L^{b \mu})^\ast}{M_{U_1}^2} \Big\vert^2,
\end{equation}

\noindent where $f_{\Upsilon}$ and $m_{\Upsilon}$ are the Upsilon decay constant and mass, respectively. The decay constant values can be extracted from the experimental branching ratio measurements of the processes $\Upsilon \to e^-e^+$. Using current data from the Particle Data Group (PDG)~\cite{PDG2020}, one obtains $f_{\Upsilon(1S)} = (659 \pm 17) \ {\rm MeV}$,  $f_{\Upsilon(2S)} = (468 \pm 27) \ {\rm MeV}$, and $f_{\Upsilon(3S)} = (405 \pm 26) \ {\rm MeV}$. In our analysis we will only take into account the UL from $\Upsilon(3S)$ that leads to the strongest bound. Belle II would be able to improve $\Upsilon(3S) \to \mu^\pm \tau^\mp$ down to $\sim 10^{-7}$~\cite{Bhattacharya:2016mcc}.

\subsection{$B \to K \tau^+ \tau^-$ and $B_s \to \tau^+ \tau^-$ decay}

We also take into account the rare $B$ decays, namely $B \to K \tau^+ \tau^-$ and $B_s \to \tau^+ \tau^-$, which are  induced via the $b \to s\tau^+ \tau^-$ transition. These decay channels have not been observed so far and the present reported bounds~\cite{Aaij:2017xqt,PDG2020} are shown in Table~\ref{TableLFV}, as well as the planned Belle II sensitivity~\cite{Kou:2018nap}. In the case of $B_s \to \tau^+ \tau^-$, an additional projected sensitivity of $\sim 5 \times 10^{-4}$ is expected at LHCb with 50 fb$^{-1}$~\cite{Albrecht:2017odf}. The branching fraction of semileptonic decay $B \to K \tau^+ \tau^-$ can be expressed by the numerical formula~\cite{Cornella:2019hct}
\begin{equation}
{\rm BR}(B \to K \tau^+ \tau^-) \simeq 1.5\times 10^{-7} + 1.4\times 10^{-3} \Big( \frac{1}{2\sqrt{2}G_F}\Big) \frac{{\rm Re}[x_L^{s \tau}(x_L^{b\tau})^\ast]}{M_{U_1}^2} + 3.5 \Big( \frac{1}{2\sqrt{2}G_F}\Big)^2 \frac{\vert x_L^{s \tau}(x_L^{b\tau})^\ast \vert^2}{M_{U_1}^4}.
\end{equation}

\noindent For the leptonic process $B_s \to \tau^+ \tau^-$, the SM branching ratio is modified as~\cite{Cornella:2019hct}
\begin{equation}
{\rm BR}(B_s^0 \to \tau^+ \tau^-) = {\rm BR}(B_s^0 \to \tau^+ \tau^-)_{\text{SM}} 
 \Bigg| 1+ \dfrac{ \pi}{\sqrt{2} G_F \alpha_{\rm em} V_{tb} V_{ts}^\ast C_{10}^{\rm SM}} \dfrac{x_L^{s \tau}(x_L^{b\tau})^\ast}{M_{U_1}^2} \Bigg|^2,
\end{equation}
\noindent where ${\rm BR}(B_s^0 \to \tau^+ \tau^-)_{\rm SM} =  (7.73 \pm 0.49) \times 10^{-7}$~\cite{Bobeth:2013uxa} and $C_{10}^{\rm SM} \simeq -4.3$.

\section{Phenomenological Analysis} 
\label{analysis}

In this section we perform a global phenomenological analysis on the parametric space of the $U_1$ vector  model (discussed in Sec.~\ref{model}) addressing the $b\to s \mu^+\mu^-$ and $b \to c \tau\bar{\nu}_\tau$ anomalies. For this analysis we define the data set \textit{All data}, which includes:
\begin{equation}
\text{All data} \in \begin{cases}
b \to c \tau\bar{\nu}_\tau \ \text{data:} \ R(D), R(D^\ast), R(J/\psi), F_L(D^*), P_\tau(D^*), {\rm BR}(B_c^- \to \tau^- \bar{\nu}_{\tau})< 10\%, R(X_c) \\
R_D^{\mu/e}, B^- \to \tau^- \bar{\nu}_{\tau} \\
b\to s \mu^+\mu^- \ \text{data:} \ (C^{bs\mu\mu}_{9} = - C^{bs\mu\mu}_{10} \ \text{solution}) \\
\text{LFV decays:} \ B \to K^{(\ast)} \mu^\pm \tau^\mp, B_s \to \mu^\pm \tau^\mp, \tau \to \mu\phi, \Upsilon(nS) \to \mu^\pm \tau^\mp \\
\text{rare $B$ decays:} \ B \to K \tau^+ \tau^-, B_s \to \tau^+ \tau^- 
\end{cases}
\end{equation}

\noindent All of the observables were previously discussed in Sec.~\ref{observables}. Let us notice that rather to provide predictions on LFV channels (as done in the recent analysis of Ref.~\cite{Angelescu:2021lln}), we take a different approach by incorporating in our study the current available ULs on their branching ratios.
Thus, we have a total of 19 observables for all data and four free-parameters $(x_L^{s\mu},x_L^{s\tau},x_L^{b\mu},x_L^{b\tau})$ of the $U_1$ LQ model to be fitted; therefore, the number of degrees of freedom ($N_{\rm dof}$) of the analysis is $N_{\rm dof} = 15$. We will include in our analysis the first measurement by LHCb on the ratio $R(\Lambda_c)$~\cite{LHCb:2022piu}, as well as the $R(\Lambda_c)$ normalization issue discussed in Ref.~\cite{Bernlochner:2022hyz} (referred to by us as $R(\Lambda_c)_{\rm Revisited}$). We also take into account the leptonic decay ratio of bottomonium meson $R_\Upsilon=R_{\Upsilon(nS)} \ (n=1,2,3)$, which includes the new BABAR measurement of  $R_{\Upsilon(3S)}$~\cite{Garcia-Duque:2021qmg}. The step by step inclusion of these new observables will increase $N_{\rm dof}$ to 16 and 21, respectively, allowing us to explore their impact on the $U_1$ model global fit. It is worth noticing that the implications of $R_\Upsilon$ are usually ignored in most of the recent (and previous) studies of the minimal $U_1$ model~\cite{Angelescu:2021lln}. As we pointed out above, NP scenarios aiming to provide an explanation to the anomalous $b \to c \tau\bar{\nu}_\tau$ data, also induce effects in the neutral-current transition $b\bar{b} \to \tau^+\tau^-$~\cite{Aloni:2017eny}. As concerns $R(\Lambda_c)$, our study is the first $U_1$ model analysis taking into account the recent LHCb result~\cite{LHCb:2022piu}. \medskip

Regarding the LHC constraints, we use in our analysis the benchmark LQ mass $M_{U_1} = 1.8 \ {\rm TeV}$, which corresponds to the lower limit obtained in Ref.~\cite{Angelescu:2021lln} from an analysis of recent LHC data based on direct and indirect high-$p_T$ searches. Moreover, we also consider the ULs on the vector LQ couplings $(x_L^{s\mu},x_L^{s\tau},x_L^{b\mu},x_L^{b\tau})$ as a function of the LQ mass, which have been obtained from the most recent LHC searches in the high-$p_T$ bins of $pp \to \ell\ell$ at 13 TeV with 140 fb${}^{-1}$~\cite{Angelescu:2021lln}. \medskip

We construct the corresponding $\chi^2$ function and obtain its minimum $(\chi^2_{\rm min})$. By considering the different data sets, in Table~\ref{fit} we show the best fit $1\sigma$ regions for the $U_1$ model parameters. To analyze the goodness of the fit, we also report the $\chi^2_{\rm min} /N_{\rm dof}$ and $p$-value of each scenario. As can be seen, all data has the largest $p$-value, thus providing the best fit. Once one incorporates $R(\Lambda_c)_{\rm LHCb}$ (or $R(\Lambda_c)_{\rm Revisited}$) into the fit, the $p$-value is reduced, without affecting (very mildly) the $1\sigma$ regions. As expected, with the inclusion of $R(\Lambda_c)_{\rm Revisited}$ we get a larger $p$-value than with $R(\Lambda_c)_{\rm LHCb}$. 
Moreover, the addition of the bottomonium observables $R_{\Upsilon}$ (mainly due to $R_{\Upsilon}(3S)$) also impact the $p$-value reducing its value, but keeping the same $1\sigma$ confidence values of LQ couplings. 
For all of the data sets considered, we observed that the best fit values  exhibit a hierarchical pattern preferring large values for the $x_L^{b\tau}\sim \mathcal{O}(1)$ coupling, while small ones for $x_L^{s\tau} \approx x_L^{b\mu} \sim \mathcal{O}(10^{-1})$ and $\vert x_L^{s\mu} \vert \sim \mathcal{O}(10^{-2})$.
In summary, after the inclusion of $R(\Lambda_c)$ and $R_{\Upsilon(3S)}$ into the analysis, the final result with the full data is that the minimal $U_1$ model provides a good fit, making it still a viable explanation of the $B$ meson anomalies. In addition, we have shown that the incorporation of these new observables generate strong tension, leading to non-trivial effects to the global fit. Future improvements on the measurements of $R(\Lambda_c)$ and $R_{\Upsilon(3S)}$ at the LHCb and Belle II experiments would help to understand the complementarity of these observables. \medskip

\begin{table}[t]
\centering
\renewcommand{\arraystretch}{1.2}
\renewcommand{\arrayrulewidth}{0.8pt}
\caption{The $1\sigma$ fit results of $U_1$ LQ couplings, $\chi^2_{\rm min} /N_{\rm dof}$, and $p-$value for different data sets. In all the cases considered, we have used the benchmark mass value of $M_{U_1} = 1.8 \ {\rm TeV}$.} \label{fit}
\begin{tabular}{ccccccc}
\hline\hline
Data set & $x_L^{s\mu}$ & $x_L^{s\tau}$  & $x_L^{b\mu}$ & $x_L^{b\tau}$ & $\chi^2_{\rm min} /N_{\rm dof}$ & $p-$value ($\%$) \\                                                                                                                        
\hline
All data & $[-0.19,0.15]$ & $[0.08,0.17]$ & $[0.13,0.18]$  & $[1.25,1.87]$ & $6.26/15$  & 97.5  \\
All data $+R(\Lambda_c)_{\rm LHCb}$ &  $[-0.17,0.14]$ & $[0.06,0.16]$ & $[0.14,0.20]$  & $[1.24,1.88]$ & $8.93/16$ & 91.6 \\ 
All data $+R(\Lambda_c)_{\rm Revisited}$ & $[-0.18,0.15]$ & $[0.07,0.16]$ & $[0.14,0.19]$  & $[1.24,1.88]$ & $7.51/16$ & 96.2 \\
All data $+R(\Lambda_c)_{\rm LHCb}+R_\Upsilon$ &  $[-0.17,0.15]$ & $[0.06,0.16]$ & $[0.14,0.19]$  & $[1.22,1.87]$ & $12.7/21$ & 85.1 \\ 
All data $+R(\Lambda_c)_{\rm Revisited}+R_\Upsilon$ & $[-0.18,0.15]$ & $[0.07,0.16]$ & $[0.14,0.19]$  & $[1.23,1.86]$ & $11.3/21$ & 91.2 \\ 
\hline\hline
\end{tabular}
\end{table} 

\begin{figure*}[!t]
\centering
\includegraphics[scale=0.40]{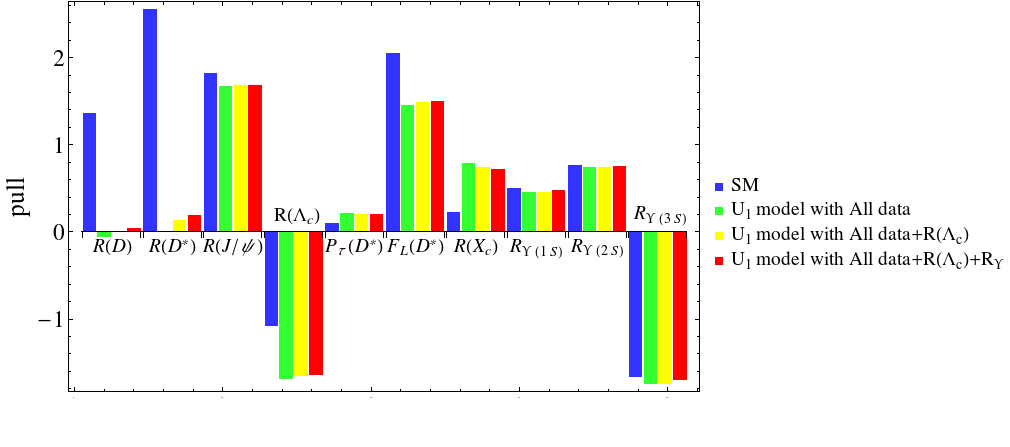}
\caption{\small Pulls of the $b\rightarrow c\tau \bar{\nu}_\tau$ and $b\bar{b} \to \tau^{+}\tau^{-}$ observables with respect to SM (blue bar) and the best-fit point of the $U_1$ model for three different data sets (green, yellow, and red bars).}
\label{Figpulls}
\end{figure*}

To complement the previous discussion we now consider Fig.~\ref{Figpulls} where the blue bars represent the pulls of the $b\rightarrow c\tau \bar{\nu}_\tau$ and $b\bar{b} \to \tau^{+}\tau^{-}$ observables with respect to SM, i.e., ${\rm pull}_i =( \mathcal{O}^{\rm exp}_i - \mathcal{O}^{\rm th}_i )/\Delta \mathcal{O}_i$, where $\mathcal{O}^{\text{exp}}_i$ stands for experimental measurement,  $\mathcal{O}^{\text{th}}_i$ its corresponding prediction by the SM (or the NP model we are considering)  and $\Delta \mathcal{O}_i= (\sigma_{\text{exp}}^2+\sigma_{\text{th}}^2)^{1/2}$ is the summation in quadrature of the experimental and theoretical uncertainties. The pull is positive when $\mathcal{O}^{\rm exp} > \mathcal{O}^{\rm th}$ (excess) and negative when $\mathcal{O}^{\rm exp} < \mathcal{O}^{\rm th}$ (deficit), such is the case of the observables $R(\Lambda_c)$ and $R_{\Upsilon(3S)}$ (see Fig.~\ref{Figpulls}).
The green, yellow, and red bars correspond to the best-fit point of the $U_1$ model for three different data sets. It is important to note that with the $U_1$ model the observables $R(D)$, $R(D^\ast)$, $R(J/\psi)$ and $F_L(D^\ast)$ decrease the tension with the experiment, particularly, there is an excellent improvement in the prediction of $R(D^{(\ast)})$.  The observables $P_\tau(D^\ast)$ and $R(X_c)$ increase the pull but it is less than one, so the model $U_1$ remains consistent with the experiment. While $R_{\Upsilon(1S)}$ and $R_{\Upsilon(2S)}$ remain (almost) unchanged. The only observables for which the pull are increased (therefore, inconsistent with the  $U_1$ model) are $R(\Lambda_c)$ and $R_{\Upsilon(3S)}$. As pointed out above, this result shows that these observables are important in the analysis of the charged-current $B$ anomalies, and should be taken into account in future analyses. \medskip

\begin{figure*}[!t]
\centering
\includegraphics[scale=0.35]{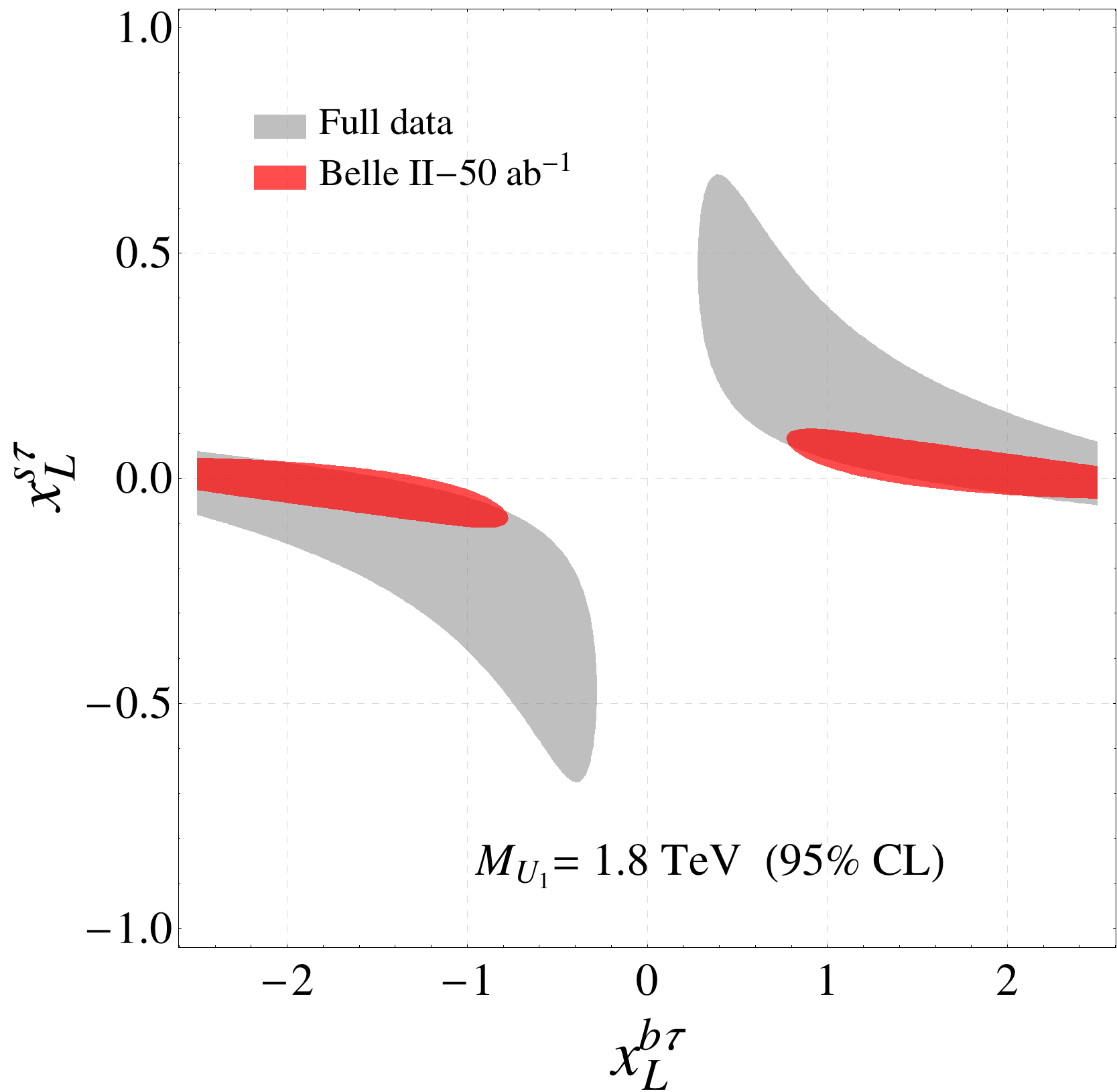} \ 
\includegraphics[scale=0.36]{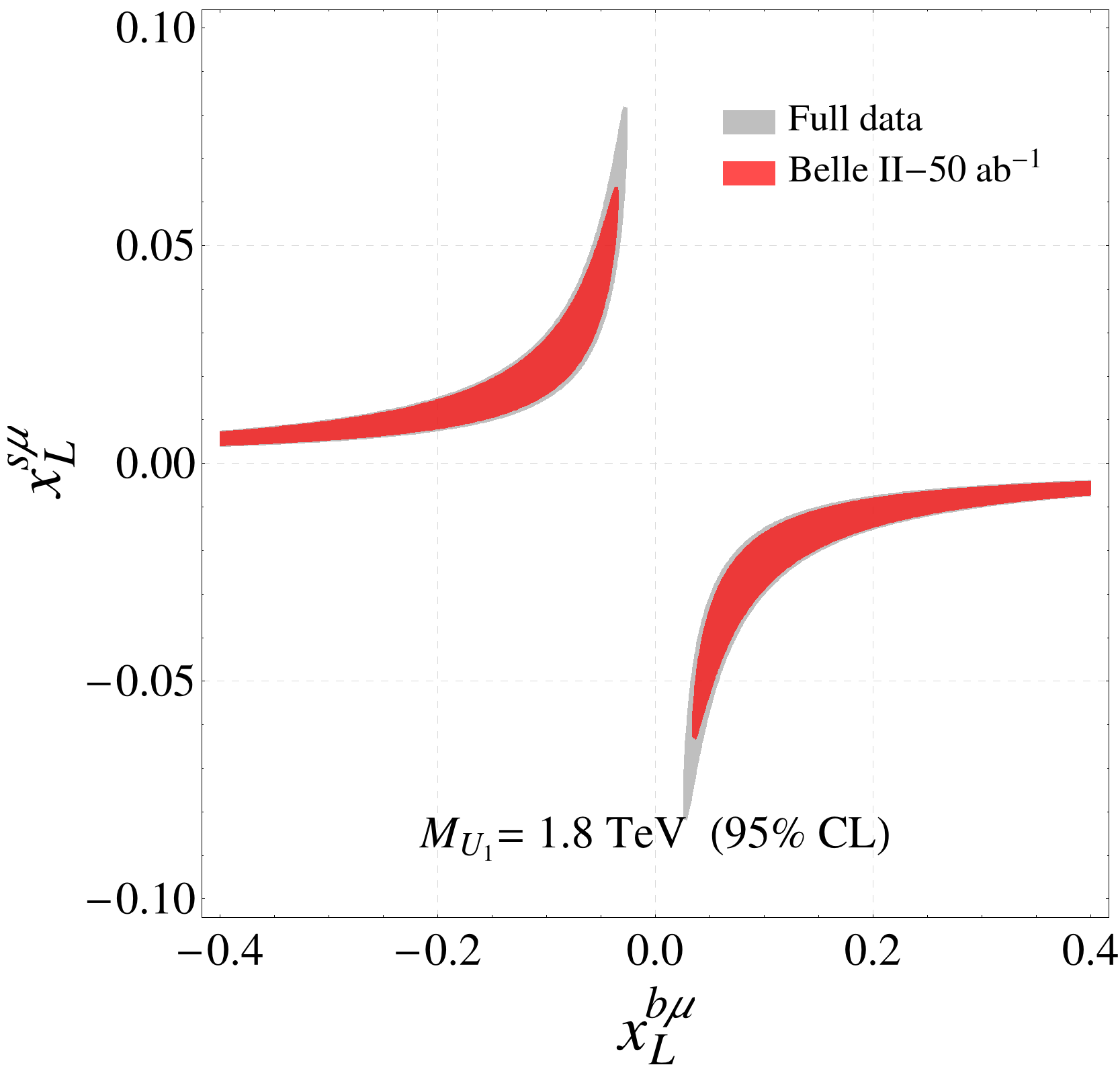}
\caption{\small The $95\%$ CL allowed regions of the planes $(x_L^{b\tau},x_L^{s\tau})$ [left] and $(x_L^{b\mu},x_L^{s\mu})$ [right] of $U_1$ vector model for $M_{U_1} = 1.8 \ {\rm TeV}$, respectively. In each plot we are marginalizing over the rest of the parameters. The gray region corresponds to the full data (all data $+R(\Lambda_c)+R_\Upsilon$). The Belle II projection for an integrated luminosity of $50 \ {\rm ab}^{-1}$ is represented by the red region.}
\label{PS}
\end{figure*}

Taking into account the previous results of full data fit (all data $+R(\Lambda_c)+R_\Upsilon$, where $R(\Lambda_c)$ stands either for $R(\Lambda_c)_{\rm LHCb}$ or $R(\Lambda_c)_{\rm Revisited}$), in Fig.~\ref{PS} we show the allowed region (gray color) at the $95\%$ confidence level (CL) of the planes $(x_L^{b\tau},x_L^{s\tau})$ [left] and $(x_L^{b\mu},x_L^{s\mu})$ [right] of $U_1$ vector model for $M_{U_1} = 1.8 \ {\rm TeV}$, respectively. Among the LFV decays, we found that the very recent UL obtained by LHCb on $B^0 \to K^{\ast 0} \mu^\pm \tau^\mp$~\cite{LHCb:2022wrs} provides the strongest bounds. In addition, for further discussion, we also include the impact on $U_1$ model from the Belle II envisaged improvements on different observables previously discussed in Sec.~\ref{observables}. These include the Belle II sensitivities (see Table~\ref{TableLFV}) on the branching fraction of LFV decays ($B^+ \to K^+ \mu^\pm \tau^\mp$, $\tau \to \mu\phi$, $\Upsilon(3S) \to \mu^\pm \tau^\mp$), rare $B$ decays ($B \to K \tau^+ \tau^-$, $B_s \to \tau^+ \tau^-$), and Belle II prospects on $R(D^{(\ast)})$~\cite{Kou:2018nap} in which $R(D^{(\ast)})$ keep the central values of Belle combination averages~\cite{Belle:2019rba} with the uncertainties improvements for an integrated luminosity of $50 \ {\rm ab}^{-1}$~\cite{Kou:2018nap}. The Belle II$-50 \ {\rm ab}^{-1}$ projection is depicted in Fig.~\ref{PS} by the red region. We obtain that the parametric space would be narrowed by $\text{Belle \ II}-50 \ {\rm ab}^{-1}$ but still allowing small room for NP. Particularly, the $(x_L^{b\tau},x_L^{s\tau})$ plane would be severely constrained  to small values of $x_L^{s\tau}$ coupling. This is a consequence of the expected improvements on $\tau \to \mu\phi$, and $B \to K \tau^+ \tau^-$. While the $(x_L^{b\tau},x_L^{s\tau})$ plane would be mainly affected by the $\tau \to \mu\phi$ decay. Thus, the searches at Belle II of these LFV and rare decays will be a matter of importance on proving the $U_1$ vector LQ explanation to the $B$ meson anomalies, as previously suggested in Ref.~\cite{Hati:2020cyn}. We want to stress that  our analysis strengthens and complements the recent phenomenological analysis of the minimal $U_1$ model presented in Ref.~\cite{Angelescu:2021lln}.

\FloatBarrier

\section{Addressing the $a_\mu=\frac{1}{2} (g-2)_\mu$ anomaly} 
\label{gminus2}

Recently, a new measurement of the anomalous magnetic moment of the muon, $a_\mu=\frac{1}{2} (g-2)_\mu$, has
been obtained by the Muon $g-2$ collaboration at Fermilab~\cite{Muong-2:2021ojo}, in excellent agreement with the previous measurement performed at BNL E821~\cite{Muong-2:2006rrc}. The combined experimental average is $a_\mu^{\rm Exp} = (116592061 \pm 41)\times 10^{-11}$, corresponding to $4.2\sigma$ deviation from the SM contribution~\cite{Muong-2:2021ojo}
\begin{equation}
\Delta a_\mu = a_\mu^{\rm Exp} - a_\mu^{\rm SM} =(251 \pm 59)\times 10^{-11}. 
\end{equation}

\noindent The $U_1$ vector LQ can contribute at the one-loop level to $(g-2)_\mu$~\cite{Altmannshofer:2020ywf,Du:2021zkq,Ban:2021tos}. In the heavy limit $M_{U_1} \gg m_\mu$, the dominant one-loop contribution can be written as~\cite{Altmannshofer:2020ywf,Ban:2021tos}
\begin{equation}
\Delta a_\mu^{U_1} = \frac{N_c m_\mu^2}{16\pi^2 M_{U_1}^2} \Big[ -(\vert x_{L}^{b\mu}\vert^2 +\vert x_{R}^{b\mu}\vert^2) \Big(\frac{4}{3} Q_b - \frac{5}{3} Q_{U_1} \Big) + 2 \ {\rm Re}[x_{L}^{b\mu} (x_{R}^{b\mu})^\ast] \frac{2m_b}{m_\mu} (Q_b - Q_{U_1}) \Big],
\end{equation}

\noindent where $m_\mu$ and $m_b$ are masses of the muon and bottom quark, respectively; $N_c=3$ is a color factor; $Q_{U_1}=+2/3$ and $Q_b = -1/3$ are the leptoquark and bottom quark electric charges, respectively. Let us notice, that in addition to the left-handed coupling $x_L^{b\mu}$ contribution, it is also necessary to add the contribution from the right-handed coupling $x_R^{b\mu}$ to explain $\Delta a_\mu$. Considering only the coupling $x_L^{b\mu}$, the effect in $\Delta a_\mu$ is small ($\Delta a_\mu \sim 10^{-12}$). The presence of both LQ couplings gives rise to an enhancement due to mass ratio $m_b/m_\mu$.  \medskip

\begin{figure*}[!t]
\centering
\includegraphics[scale=0.35]{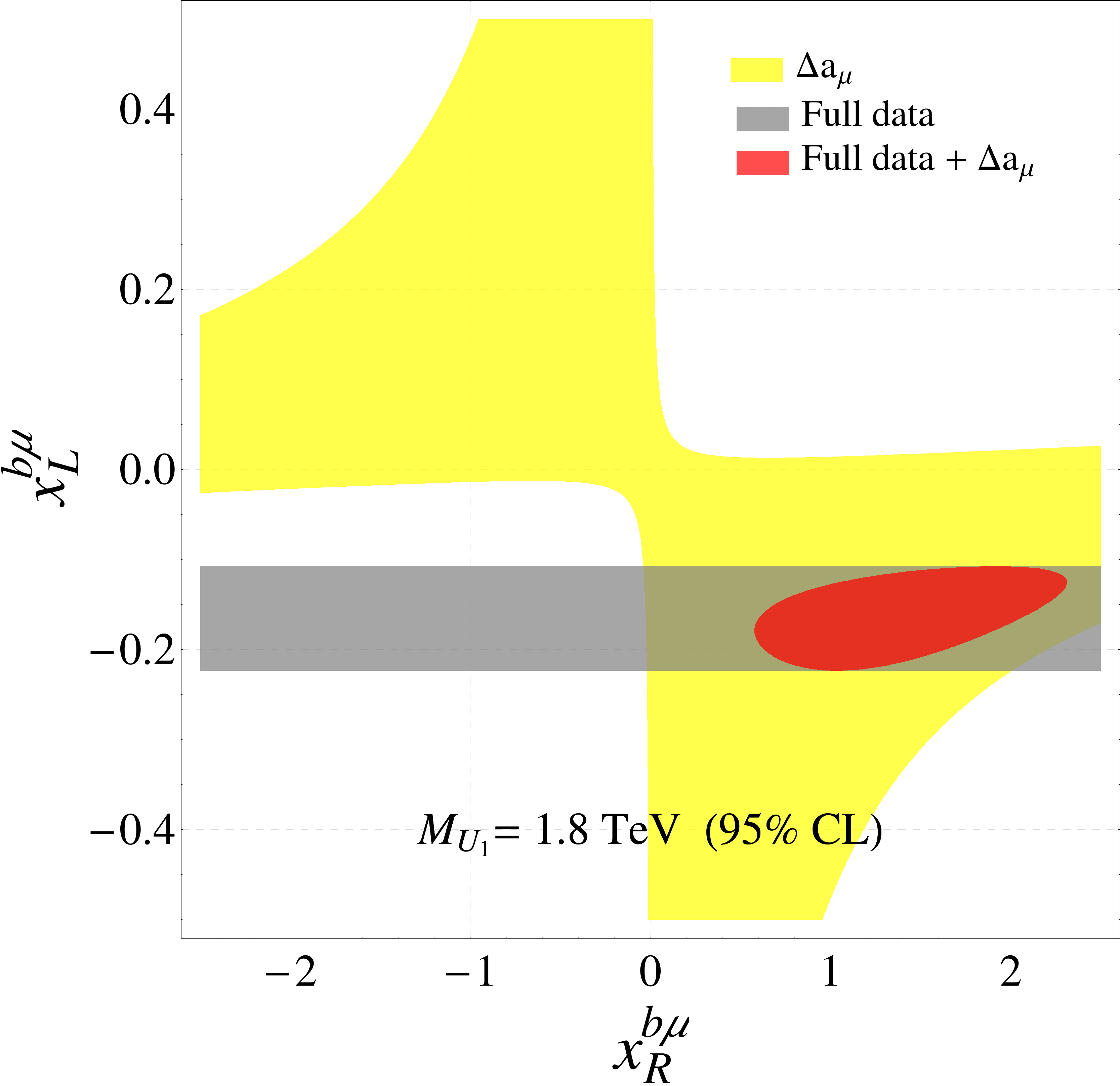}
\caption{\small The $95\%$ CL allowed parameter space in the ($x_R^{b\mu},x_L^{b\mu}$) plane for $\Delta a_\mu$ (yellow region), full data (gray region), and full data+$\Delta a_\mu$ $x_L^{b\mu}$ (red region), respectively.}
\label{Fig2}
\end{figure*}


In the following, we will extend our previous phenomenological analysis of the minimal $U_1$ model to include $\Delta a_\mu$. For such a purpose, we add $\Delta a_\mu$ to full data (= all data $+R(\Lambda_c)_{\rm LHCb}+R_\Upsilon$) and carry out a $\chi^2$ analysis. For this global fit, we now have 20 observables and five free-parameters $(x_L^{s\mu},x_L^{s\tau},x_L^{b\mu},x_L^{b\tau},x_R^{b\mu})$. We find the following best fit points and $1\sigma$ solutions for the LQ couplings,
\begin{eqnarray}
x_L^{s\mu} =0.014 \  & \to & \ [-0.14,0.17], \\
x_L^{s\tau} =0.11 \  & \to & \ [0.06,0.16], \\
x_L^{b\mu} =-0.16 \  & \to & \ [-0.19,-0.14], \\
x_L^{b\tau} =1.55 \  & \to & \ [1.22,1.88], \\
x_R^{b\mu} =1.35 \  & \to & \ [1.05,1.64],
\end{eqnarray}
  
\noindent respectively. For this analysis we get a good fit of data with $\chi^2_{\rm min} /N_{\rm dof} = 0.60$ and $p$-value $=91.2 \%$. In comparison with our previous analysis (see Table~\ref{fit}), we notice that the $1\sigma$ solutions are slightly modified and there is  a change in the individual signs of $x_L^{s\mu}$ and $x_L^{b\mu}$ (their product has to be negative to fulfill the $C^{bs\mu\mu}_{9} = - C^{bs\mu\mu}_{10}$ solution). One important remark is that the RH coupling must have large values, but below the perturbative regime ($ \lesssim \sqrt{4\pi}$).  
Additionally, for a mass of $M_{U_1} = 1.8$ TeV in Fig.~\ref{Fig2} we present the parameter space in the ($x_R^{b\mu},x_L^{b\mu}$) plane, where the regions in yellow, gray, and red are associated with the allowed regions by $\Delta a_\mu$, full data, and full data+$\Delta a_\mu$, respectively, at the $95\%$ CL.
Therefore, our analysis shows that the minimal $U_1$ model can be economically extended with (large) right-handed coupling $x_R^{b\mu}$ to simultaneously address the anomalies in $a_\mu$ and $b \to c \tau\bar{\nu}_\tau$ and $b\to s \mu^+\mu^-$ data.


\section{Concluding remarks} \label{Conclusion}

The recent measurements of the LHCb on the LFU ratio $R(\Lambda_c) = {\rm BR}(\Lambda_b \to \Lambda_c\tau\bar{\nu}_\tau)/{\rm BR}(\Lambda_b \to \Lambda_c\mu\bar{\nu}_\mu)$ and the BABAR on the leptonic decay ratio of the Upsilon meson $\Upsilon(3S)$, $R_{\Upsilon(3S)} = {\rm BR}(\Upsilon(3S) \to \tau^+\tau^-)/{\rm BR}(\Upsilon(3S) \to \mu^+\mu^-)$, are suppressed in comparison with their corresponding SM predictions. To the light of these new measurements, we have reanalized the combined explanation of the semileptonic $B$ meson anomalies within the singlet vector LQ model (the so-called minimal $U_1$ model). 
For the $b \to c \tau \bar{\nu}_\tau$ data, we have included $R(D^{(\ast)})$, $R(J/\psi)$, $F_L(D^*)$, $P_\tau(D^*)$, ${\rm BR}(B_c^- \to \tau^- \bar{\nu}_{\tau})< 10\%$, and $R(X_c)$ observables, as well as $R_D^{\mu/e}$ and $B\to \tau\bar{\nu}_\tau$. While for the $b \to s \mu^+\mu^-$ data, we used the $C^{bs\mu\mu}_{9} = - C^{bs\mu\mu}_{10}$ solution preferred by the global fit analyses. The minimal $U_1$ model is also constrained by a number of tree-level induced processes such as, LFV decays ($B\to K^{(\ast)} \mu^\pm \tau^\mp$, $B_s \to \mu^\pm \tau^\mp$, $\tau \to \mu\phi$, $\Upsilon(nS) \to \mu^\pm \tau^\mp$), rare $B$ decays ($B \to K \tau^+ \tau^-, B_s \to \tau^+ \tau^-$), and bottomonium ratios $R_{\Upsilon(nS)}$; which we have properly taken into account. In addition, we have incorporated in our analysis the LHC constraints and the expected improvements on different flavor processes at Belle II for an integrated luminosity of $50 \ {\rm ab}^{-1}$.\medskip

We carried out a global fit of the phenomenology (allowed parametric space) of the relevant flavor-dependent couplings between $U_1$ and left-handed SM fermions $(x_L^{s\mu},x_L^{s\tau},x_L^{b\mu},x_L^{b\tau})$. For a benchmark mass value of $M_{U_1} = 1.8$ TeV, the main finding of our study is that the minimal $U_1$ model is still one of the simplest combined explanation of the $B$ meson anomalies, providing a good fit of the current full data. 
Nevertheless, our results showed that the inclusion of the new observables $R(\Lambda_c)$ and $R_{\Upsilon(3S)}$ generates a non-trivial tension into the global fit, yielding to a worsening of the $p-$value (goodness of the fit); therefore, these observables cannot be individually accommodated within the minimal $U_1$ model. Future improvements on the $R(\Lambda_c)$ and $R_{\Upsilon(3S)}$ measurements at the LHCb and Belle II experiments would help to clarify this situation. On the other hand, regarding the Belle II perspectives, we have also found that the parametric space would be narrowed as a result of the expected improvements on $\tau \to \mu\phi$ and $B \to K \tau^+ \tau^-$ decays. Our study thus confirms the potential of Belle II to provide a complementary test of the $U_1$ model.  \medskip

Finally, we have shown that the (long-standing) current tension in the anomalous magnetic moment of the muon $a_\mu$ can be also addressed by economically extending the minimal $U_1$ model with the addition of the right-handed bottom-muon coupling ($x_R^{b\mu} \neq 0$) with large values. As a consequence, the $B$ meson anomalies ($b \to c \tau\bar{\nu}_\tau$ and $b\to s \mu^+\mu^-$ data) and $a_\mu$ can be simultaneously explained within this singlet vector LQ model.


\acknowledgments
J. H. Mu\~{n}oz is grateful to Vicerrectoria de Investigaciones of Universidad del Tolima for financial support of Project No. 290130517. The work of N. Quintero has been financially supported by Universidad Santiago de Cali. E. Rojas acknowledges financial support from the “Vicerrectoría de Investigaciones e Interacción Social VIIS de la Universidad de Nariño,” Projects No. 1928 and No. 2172.


\end{document}